\documentclass[aps,prl,preprint,showpacs]{revtex4}

\usepackage{amsfonts,amsmath,amssymb,graphicx}
\usepackage{amsfonts}
\usepackage{amsmath}
\usepackage{amssymb}
\usepackage{graphicx}

\def\kbar{\protect\@kbar}
\def\@kbar{\relax \bgroup
\def\@tempa{\hbox{\raise.73\ht0
\hbox to0pt{\kern.25\wd0\vrule width.5\wd0 height.1pt
depth.1pt\hss}\box0}}\mathchoice{\setbox0\hbox{$\displaystyle
k$}\@tempa}{\setbox0\hbox{$\textstyle
k$}\@tempa}{\setbox0\hbox{$\scriptstyle
k$}\@tempa}{\setbox0\hbox{$\scriptscriptstyle k$}\@tempa}\egroup}

\begin{document}

\title{\textbf{Statistical Approach to Quantum Chaotic Ratchets}}
\author{Itzhack Dana}
\affiliation{Minerva Center and Department of Physics, Bar-Ilan University, 
Ramat-Gan 52900, Israel}

\begin{abstract}
The quantum ratchet effect in fully chaotic systems is approached by
studying, for the first time, \emph{statistical} properties of the ratchet
current over well-defined sets of initial states. Natural initial states in
a semiclassical regime are those that are \emph{phase-space uniform} with
the \emph{maximal possible} resolution of one Planck cell. General arguments
in this regime, for quantum-resonance values of a scaled Planck constant $\hbar$, predict that the distribution of the current over all such
states is a zero-mean Gaussian with variance $\sim D\hbar ^{2}/(2\pi ^{2})$,
where $D$ is the chaotic-diffusion coefficient. This prediction is well supported by extensive numerical evidence. The average strength of the effect, measured by the variance above, is \emph{significantly larger} than that for the usual momentum states and other states. Such strong effects should be  experimentally observable.

\end{abstract}

\pacs{05.45.Mt, 05.45.Ac, 03.65.-w, 05.60.Gg}
\maketitle

\begin{center}
{\bf I. INTRODUCTION}
\end{center}

Understanding quantum transport in generic Hamiltonian systems, which are
classically nonintegrable and exhibit chaos, is a problem of both
fundamental and practical importance. The study of simple model systems
have led to the discovery of a variety of quantum-transport phenomena
\cite{qc,er,ht,khm,ls,d,d0,d1,kh,hr,qr,qr1,qr2,qr3,qr4,qr5,qr6,qr7}, several
of which have been observed in atom-optics experiments
\cite{er,ht,qr5,qr6,qr7} and allow to control the quantum motion of cold
atoms or Bose-Einstein condensates in different ways. Recently, classical
and quantum Hamiltonian \emph{\textquotedblleft ratchets\textquotedblright
} have started to attract a considerable interest, both theoretically
\cite{hr,qr,qr1,qr2,qr3,qr4} and experimentally \cite{qr5,qr6,qr7}.
Ratchets are usually conceived as spatially periodic systems with noise
and dissipation in which a directed current of particles can emerge from
an unbiased (zero-mean) external force due to some spatial/temporal
asymmetry \cite{rat}. In classical Hamiltonian ratchets \cite{hr},
dissipation is absent and noise is replaced by deterministic chaos. A
basic general result for Hamiltonian dynamics under an unbiased force is
that the average current of an initially \emph{uniform} ensemble of
particles in phase space is zero \cite{hr}. As a consequence, a completely
chaotic system carries essentially no ratchet current. On the other hand,
the corresponding quantized system can feature significant ratchet effects
\cite{qr,qr1,qr2,qr3,qr4,qr5,qr6,qr7}. An important problem is to
understand the nature of these full-chaos quantum effects in a
semiclassical regime, in particular how precisely they vanish, as
expected, in the classical limit. All the studies of quantum chaotic
ratchets until now have mainly focused on the impact of several kinds of
asymmetries on the quantum directed current from a \emph{fixed} initial
state. It is, however, well established that the current is
\emph{sensitive} to the initial state \cite{hr,qr1,qr2,qr4,qr6,qr7} and
this sensitivity is expected to be especially high in a semiclassical
full-chaos regime, reflecting the exponential sensitivity of chaotic
motion to initial conditions. Thus, to get a comprehensive understanding
of the quantum ratchet effect, it is necessary to adopt a more
\emph{global} approach, \emph{not} limited to a single initial
state.\newline

In this paper, the semiclassical full-chaos regime of quantum ratchets is
approached by studying, for the first time, \emph{statistical} properties of
the current over sets of initial states with well-defined natural
characteristics. The systems considered are generalizations of the
paradigmatic kicked Harper models (KHMs) \cite{qc,khm,ls,d,d0,d1,kh,qr3,qr4}, 
with Hamiltonian
\begin{equation}
\hat{H}=L\cos (\hat{p})+KV(\hat{x})\sum_{t=-\infty }^{\infty }\delta
(t^{\prime }-t),  \label{khm}
\end{equation}
where $L$ and $K$ are parameters, $\hat{x}$ and $\hat{p}$ are scaled
position and momentum operators, $V(\hat{x})$ is a general $2\pi
$-periodic potential, $t^{\prime }$ is time, and $t$ is the integer time
labeling the kicks. Generalized KHMs such as (\ref{khm}) describe several
realistic systems \cite{d,d0,kh,qr4}, in particular they are exactly
related \cite{d,d0} to kicked harmonic oscillators, which are
experimentally realizable by atom-optics methods \cite{ht}, and to kicked
charges in a magnetic field \cite{d0}. Recently \cite{qr3}, the systems
(\ref{khm}) were shown to exhibit generically a significant and robust
quantum momentum current (ratchet acceleration) under full-chaos conditions. 
The initial state
was chosen, as in other works, as a zero-momentum state. In our
statistical approach, we identify natural initial states for the
semiclassical regime as those that are analogous as much as possible to a
phase-space uniform ensemble, for which classical ratchet effects are
totally absent. These are states which are uniform in \emph{phase space}
with the \emph{maximal possible} resolution of one Planck cell. Such
uniformity is not featured by a momentum state which is uniform in
position but is infinitely localized in momentum.\newline 

Assuming quantum-resonance values of a scaled Planck constant $\hbar =[\hat{x},\hat{p}]/i$ in the semiclassical regime, we derive an estimate for the distribution of the quantum momentum current $I$ over maximally uniform initial states: This distribution is a Gaussian with mean $\left\langle I\right\rangle =0$ and variance $(\Delta I)^{2}=\left\langle I^{2}\right\rangle \sim D\hbar
^{2}/(2\pi ^{2})$, where $D$ is the chaotic-diffusion coefficient. A good agreement is found between this estimate and extensive numerical results using an exact formula for $I$ which we also derive. Examples are shown in Fig. 1 and will be discussed further in Sec. IV. The average strength of the effect, measured by
the variance above, is found to be \emph{significantly larger} than that
for the usual momentum states and other states exhibiting also zero-mean Gaussian current distributions (see the insets of Fig. 1 and Sec. IV). Our results should be experimentally observable to some extent using states approximating the maximally uniform states.\newline
\begin{figure}[tbp]
\includegraphics[width=7.5cm]{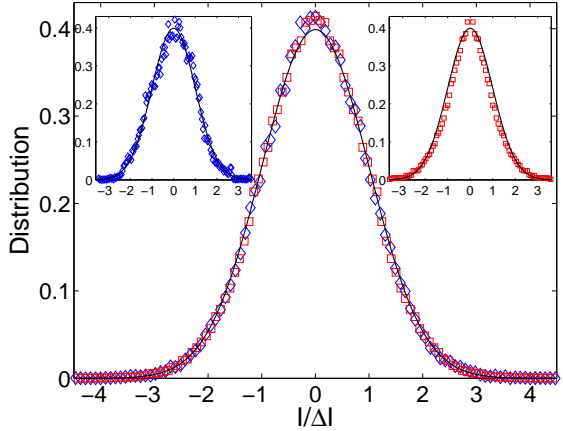}
\caption{(Color online) Distributions of the normalized quantum momentum current $I/\Delta I$ over maximally uniform states for $\hbar =2\protect\pi /121$ in two
extreme cases of fully chaotic systems (\protect\ref{khm}) with $K=15$: The
symmetric case \textquotedblleft $S$\textquotedblright\ with $V(x)=\cos (x)$
and $L=K$ (red squares, $\Delta I=0.086$) and the strongly asymmetric case
\textquotedblleft $A$\textquotedblright\ with $V(x)=\cos (x)+\sin (2x)$ and $
L=K/2$ (blue diamonds, $\Delta I=0.214$); the latter case was studied in Ref. 
\protect\cite{qr3} for a zero-momentum initial state. The origin of ratchet
currents in case $S$ is explained in Sec. IV. The insets show the
distribution of $I/\Delta I$ over momentum states in case $A$ with $K=20$
(left inset, $\Delta I=0.043$) and over low-order approximations of the maximally uniform states in case $S$\ with $K=15$ (right inset, $\Delta I=0.026$), see Sec. IV for more details. The solid line in all plots is a zero-mean Gaussian with variance 1.}
\label{fig1}
\end{figure}

The paper is organized as follows. In Sec. II, we define maximally uniform states in phase space. The main result, i.e., an estimate of the momentum-current distribution over these states in a semiclassical full-chaos regime, is derived in Sec. III. Numerical evidence for this result is provided in Sec. IV, where we also study the momentum-current distributions of states which approximate the maximally uniform states; momentum states are the crudest approximating states. Conclusions are presented in Sec. V, where we briefly mention possible experimental realizations of the strong quantum-ratchet effects predicted. Detailed derivations of some exact results are given in the Appendix.\newline
 
\begin{center}
{\bf II. MAXIMALLY UNIFORM STATES IN PHASE SPACE}
\end{center}

Maximally uniform states in phase space are defined on the basis of the
translation operators $\hat{T}_{x}(a)=\exp (i\hat{p}a/\hbar )$ and
$\hat{T} _{p}(b)=\exp (-i\hat{x}b/\hbar )$ shifting $\hat{x}$ and
$\hat{p}$ by $a$ and $b$, respectively. A state $\left\vert \psi
\right\rangle $ is uniform on the phase-space lattice with unit cell
formed by $a$ and $b$ if it is invariant under application of
$\hat{T}_{x}(a)$\ and $\hat{T}_{p}(b)$ up to constant phase factors:
$\hat{T}_{x}(a)\left\vert \psi \right\rangle =\exp (i\alpha )\left\vert
\psi \right\rangle $ and $\hat{T}_{p}(b)\left\vert \psi \right\rangle
=\exp (-i\beta )\left\vert \psi \right\rangle $. This means that
$\left\vert \psi \right\rangle $ is a simultaneous eigenstate of $\hat{T
}_{x}(a)$\ and $\hat{T}_{p}(b)$, so that these operators must commute.
Using $\hat{T}_{x}(a)\hat{T}_{p}(b)=\exp (-iab/\hbar
)\hat{T}_{p}(b)\hat{T}_{x}(a)$, we see that
$[\hat{T}_{x}(a),\hat{T}_{p}(b)]=0$ only if $ab$ is a multiple of $h=2\pi
\hbar $. Maximally uniform states $\left\vert \psi \right\rangle $
correspond to the smallest unit cell, i.e., the Planck cell with area $ab=h$.
In this case, one can easily check that the position representation of the
eigenstates $\left\vert \psi \right\rangle $ is explicitly given by
\cite{kqz}
\begin{equation}
\langle x\left\vert \psi _{\mathbf{w}}\right\rangle =\frac{1}{\sqrt{b}}\sum_{n=-\infty}^{\infty}e^{2\pi
inw_{2}/b}\delta (x-w_{1}-na),  \label{kq}
\end{equation}
where $\mathbf{w}=(w_{1},w_{2})$ range in the Planck cell, $0\leq
w_{1}<a$, $ 0\leq w_{2}<b$, and specify the phases $\alpha $ and $\beta $
above: $\alpha =w_{2}a/\hbar $ and $\beta =w_{1}b/\hbar $. The states (\ref{kq}) for all $\mathbf{w}$ form a complete and orthonormal set \cite{kqz}. Simple choices of $(a,b)$ can be made by observing that the one-period evolution operator for (\ref{khm}), $ \hat{U}=\exp [-L\cos (\hat{p})/\hbar ]\exp
[-KV(\hat{x})/\hbar ]$, is $2\pi$-periodic in both $(\hat{x},\hat{p})$.
Thus, the torus $T^{2}$: $0\leq x,p<2\pi $ is a reduced phase space for
the system. For simplicity, we shall assume from now on that there are
precisely an integer number $N$ of Planck cells within $T^{2}$, choosing
$a=2\pi /N$ and $b=2\pi $. Then, $ \hbar =ab/(2\pi )=2\pi /N$, so that the
semiclassical regime $\hbar \ll 1$ corresponds to $N\gg 1$. As it is well
known \cite{khm,d1}, the values $2\pi /N$ of $\hbar$ are those for which a
classical-quantum correspondence can be most easily established for
systems describable on a phase-space torus. These values correspond to the
main quantum resonances in the semiclassical regime.\newline

\begin{center}
{\bf III. SEMICLASSICAL ESTIMATE OF\\ THE MOMENTUM-CURRENT DISTRIBUTION}
\end{center}

The momentum-current operator $\hat{I}$ can be formally defined in the Heisenberg picture as $\hat{I}=\lim_{t\rightarrow\infty}\hat{U}^{-t}\,\hat{p}\,\hat{U}^{t}/t$, for integer time $t$. Then, the momentum current $I({\mathbf{w}})$ for initial state (\ref{kq}) is the expectation value of $\hat{I}$ in (\ref{kq}). More precisely, we show in the Appendix that in the basis of states (\ref{kq}) $\hat{I}$ is essentially represented by $\langle\psi _{\mathbf{w}}\vert\hat{I}\vert\psi _{\mathbf{w'}}\rangle =I({\mathbf{w}})\delta(w_1-w'_1)\delta(w_2-w'_2)$, where $I({\mathbf{w}})$ is given by the explicit exact formula:
\begin{equation}
I(\mathbf{w})=-\hbar \sum_{j=1}^{N}|\phi _{j}(0;\mathbf{w})|^{2}\frac{
\partial E_{j}(\mathbf{w})}{\partial w_{1}}.  \label{Ipe}
\end{equation}
Here $\phi _{j}(0;\mathbf{w})$, $j=1,\dots ,N$, are coefficients appearing
in expressions connecting states (\ref{kq}) with the $N$ quasienergy
(Floquet) eigenstates $\left\vert \Psi _{j,\mathbf{w}}\right\rangle $ of the
evolution operator $\hat{U}$ for $\hbar =2\pi /N$ and $E_{j}(\mathbf{w})$
are the corresponding quasienergy levels.\newline 

We now derive from general arguments the following estimate for the distribution $\Gamma(I)$ of $I(\mathbf{w})$ over $\mathbf{w}$ in a semiclassical full-chaos regime:
 \begin{equation}
\Gamma (I)\sim \frac{1}{\sqrt{2\pi }\Delta I}\exp \left[ -\frac{I^{2}}{
2(\Delta I)^{2}}\right] ,\ \ \ (\Delta I)^{2}\sim \frac{2D}{N^{2}}=\frac{
D\hbar ^{2}}{2\pi ^{2}},  \label{DI}
\end{equation}
where $D$ is the chaotic-diffusion coefficient. To derive (\ref{DI}), we first identify natural classical analogs of the states (\ref{kq}). To this end, let us calculate the momentum representation $\langle p\left\vert \psi _{\mathbf{w}
}\right\rangle =\int_{-\infty }^{\infty }\exp (-ipx/\hbar )\langle
x\left\vert \psi _{\mathbf{w}}\right\rangle dx$ of $\left\vert \psi _{
\mathbf{w}}\right\rangle $. One has, up to an irrelevant constant factor,
\begin{equation}
\langle p\left\vert \psi _{\mathbf{w}}\right\rangle =\sum_{n=-\infty}^{\infty}e^{-2\pi
inw_{1}/a}\delta (p-w_{2}-nb).  \label{kqp}
\end{equation}
It is clear from the delta combs (\ref{kq}) and (\ref{kqp}) that $\left\vert
\psi _{\mathbf{w}}\right\rangle $ is associated with the phase-space lattice 
$(x,p)=\mathbf{w}+\mathbf{z}(\mathbf{n})$, where $\mathbf{z}(\mathbf{n}
)=(n_{1}a,n_{2}b)=(2\pi n_{1}/N,2\pi n_{2})$ for all integers $\mathbf{n}
=(n_{1},n_{2})$. In fact, one can easily show that the Husimi distribution
of (\ref{kq}) is peaked on every point of the lattice $\mathbf{w}+\mathbf{z}(
\mathbf{n})$. This lattice, viewed as an initial phase-space ensemble, is
thus a classical analogue of $\left\vert \psi _{\mathbf{w}}\right\rangle $.
Next, consider the classical one-period map $M$ for the systems (\ref{khm}): 
$p_{t+1}=p_{t}+Kf(x_{t})$, $x_{t+1}=x_{t}-L\sin (p_{t+1})$, where 
$f(x)=-dV/dx $. For initial conditions $\mathbf{z}_{0}=(x_{0},p_{0})$, the
classical momentum current in $t$ iterations is $I_{\mathrm{c},t}(\mathbf{z}
_{0})=\Delta p_{t}(\mathbf{z}_{0})/t$, where $\Delta p_{t}(\mathbf{z}
_{0})=p_{t}-p_{0}$. Since the map $M$ is clearly $2\pi$-periodic in both $x$
and $p$, one can restrict $\mathbf{z}_{0}$ to the phase-space torus $T^{2}$:
$0\leq x,p<2\pi $. Accordingly, the initial ensemble $\mathbf{w}+\mathbf{z}(\mathbf{n})$
with $\mathbf{z}(\mathbf{n})=(2\pi n_{1}/N,2\pi n_{2})$ can be restricted to
a finite lattice of $N$ points in $T^{2}$ with $n_{1}=0,\dots ,N-1$ and $
n_{2}=0$. The classical analog of the quantum current $I(\mathbf{w})$ is
the average $\bar{I}_{\mathrm{c},t}(\mathbf{w})$ of $I_{\mathrm{c},t}(
\mathbf{z}_{0})=\Delta p_{t}(\mathbf{z}_{0})/t$, with $\mathbf{z}_{0}=
\mathbf{w}+\mathbf{z}(\mathbf{n})$, over this finite lattice:
\begin{equation}
\bar{I}_{\mathrm{c},t}(\mathbf{w})=\frac{1}{Nt}\sum_{\mathbf{n}}\Delta p_{t}[
\mathbf{w}+\mathbf{z}(\mathbf{n})],  \label{ai}
\end{equation}
for some time $t$ to be specified below. Now, under strong-chaos conditions
(large $K$ and $L$) and for sufficiently large $t$, each of the $N$
quantities $\Delta p_{t}[\mathbf{w}+\mathbf{z}(\mathbf{n})]$ in (\ref{ai})
is expected to behave diffusively, i.e., to be distributed over $\mathbf{w}$
approximately as a Gaussian with mean $\left\langle \Delta
p_{t}\right\rangle =0$ and variance $\left\langle \left( \Delta p_{t}\right)
^{2}\right\rangle \approx 2Dt$. Since these $N$ quantities are associated
with $N$ different initial points $\mathbf{w}+\mathbf{z}(\mathbf{n})$ in the
chaotic region, they should behave as independent (uncorrelated) random
variables. It then follows from the central limit theorem that for large
enough $N$ the average current (\ref{ai}) is distributed over $\mathbf{w}$
as a Gaussian with $\left\langle \bar{I}_{\mathrm{c},t}\right\rangle =0$ and 
$\left\langle \bar{I}_{\mathrm{c},t}^{2}\right\rangle \approx N\left\langle
\left( \Delta p_{t}\right) ^{2}\right\rangle /(Nt)^{2}\approx 2D/(Nt)$. This
shows how $\left\langle \bar{I}_{\mathrm{c},t}^{2}\right\rangle $ decays to
zero as $t\rightarrow \infty $, when chaotic orbits explore ergodically all
the phase space. Since there are precisely $N$ Planck cells in $T^{2}$, a
typical such orbit will explore phase space, after a time\ $t\sim N$, up to
the maximal quantum resolution of one Planck cell. Then, the
distribution of $I(\mathbf{w})$ over $\mathbf{w}$ is expected to be
approximately the same as that of the classical currents (\ref{ai}) for $t=N$,
i.e., a zero-mean Gaussian with variance $(\Delta I)^{2}=\left\langle
I^{2}\right\rangle \sim 2D/N^{2}$; this is Eq. (\ref{DI}).\newline

\begin{center}
{\bf IV. NUMERICAL EVIDENCE AND APPROXIMATING STATES}
\end{center}

In this section, we provide numerical evidence for the semiclassical estimate (\ref{DI}) using the exact formula (\ref{Ipe}) and study the momentum-current distributions for states which approximate the maximally uniform states; at the lowest order of approximation, the approximating states are just momentum states. First, the distribution $\Gamma (I)$ was calculated using (\ref{Ipe}) for several potentials $V(x)$ and many large values of $K$ and $L$ in (\ref{khm}). A good agreement was generally found between $\Gamma (I)$ and a zero-mean Gaussian distribution for sufficiently small $\hbar $. As representative examples, Fig. 1 shows distributions of $I/\Delta I$ in two extreme cases \textquotedblleft $S$\textquotedblright\ and \textquotedblleft $A$\textquotedblright\ described in the caption. For the assumed quantum-resonance values $2\pi /N$ of $\hbar$, the origin of ratchet currents in the symmetric case $S$  is the same as that already established in recent theoretical \cite{qr2} and experimental \cite{qr6,qr7} works on quantum-resonance ratchets: This is a relative asymmetry caused by the
non-coincidence of the symmetry centers of a symmetric potential with those
of a symmetric initial state. The potential $V(x)=\cos (x)$ in case $S$ has
symmetry centers at $x=0,\pi $ while the state (\ref{kq}) has them at $
x=w_{1},\ w_{1}+a/2$. Thus, for generic values of $w_{1}$, $I(\mathbf{w}
)\neq 0$.\newline 

Natural approximations of the maximally uniform states, denoted in what follows by $\left\vert\psi_{\mathbf{w}}^{(B)}\right\rangle$ for integer $B$, correspond to truncations of the infinite sum in (\ref{kqp}):
\begin{equation}
\langle p\left\vert\psi_{\mathbf{w}}^{(B)}\right\rangle =\sum_{n=-B}^B e^{-2\pi
inw_{1}/a}\delta (p-w_{2}-nb).  \label{kqB}
\end{equation}
The states (\ref{kqB}) are superpositions of the $2B+1$ momentum states $\left\vert p=w_{2}+nb\right\rangle$, $|n|\leq B$, and should be experimentally realizable (see next section). In particular, $\left\vert\psi _{\mathbf{w}}^{(0)}\right\rangle$ are just momentum states with $p=w_{2}$. We denote by $I_{B}(\mathbf{w})$ the momentum current for initial state (\ref{kqB}) and by $(\Delta I_{B})^2$ the corresponding variance over $\mathbf{w}$. The left inset of Fig. 1 shows the distribution of $I_{B}(\mathbf{w})/\Delta I_{B}$ over $\mathbf{w}$ for $B=0$ (momentum states) in case $A$ while the right inset shows it for $B=2$ in case $S$; the currents for momentum states vanish in case $S$. The fact that these distributions are again approximately zero-mean Gaussians could be expected from the simple relation (\ref{IBw}) between $I_{B}(\mathbf{w)}$ and $I(\mathbf{w})$ derived in the Appendix. Actually, we show in the Appendix that this relation leads to the exact result
\begin{equation}\label{ineq}
(\Delta I_{B})^2=\left\langle I_B^{2}\right\rangle \leq (\Delta I)^2 =\left\langle I^{2}\right\rangle .
\end{equation}

The quantities $\Delta I$ and $\Delta I_B$ were extensively studied as functions of several parameters. Consider the naturally normalized variance $R\equiv
N^{2}(\Delta I)^{2}/(2D_{\mathrm{ql}})$, where
$D_{\mathrm{ql}}=K^{2}\int_{0}^{2\pi }[f(x)]^{2}dx/2$ is the
``quasilinear" value of the diffusion coefficient $D$, obtained from the
KHM map $M$ above by neglecting all the force-force correlations
$C_t=\langle f(x_0)f(x_t)\rangle$, $t\neq 0$; for sufficiently strong
chaos, $D$ is very close to $D_{\mathrm{ql}}$. The semiclassical estimate for the variance in Eq. (\ref{DI}) would imply that
$R\approx D/D_{\mathrm{ql}}$. Indeed, Fig. 2 shows a reasonably good agreement
between $R$ and $D/D_{\mathrm{ql}}$ versus $K$ in both cases $S$ and $A$.
Discrepancies arise mainly around peaks of $D/D_{\mathrm{ql}}$, especially
the peak near $K\approx 6.5$ in case $S$, due to a small accelerator-mode
island. Thus, for general large $K$ with $R\approx 1$, $\Delta I$
increases almost linearly with $K$ like $\sqrt{D_{\mathrm{ql}}}$.\newline 

Fig. 3 shows loglog plots of $\Delta I$ versus $N$ in cases $S$ and $A$. The
results agree very well with the $N^{-1}$ behavior predicted by
(\ref{DI}). Fig. 4 shows plots of $\Delta I_{B}/\Delta I$ versus $B$ in
the two cases. We see that $\Delta I_{B}$ is always smaller than $\Delta I$, in accordance with the exact inequality (\ref{ineq}), and approaches monotonically $\Delta I$ as the order $B$ of approximation increases. For small $B$, $\Delta I_{B}$ is significantly smaller than $\Delta I$ and attains its minimal value at $B=0$, corresponding to momentum states ($\Delta I_{0}=0$ in case $S$). Within the limited domain of $N$ we could study numerically, $\Delta I_{B}$ appears to decay with $N$ like $N^{\mu}$, where $\mu$ ranges between $-0.9$ to $-1.1$ with an error not smaller than $\pm 0.03$.\newline
\begin{figure}[tbp]
\includegraphics[width=7.5cm]{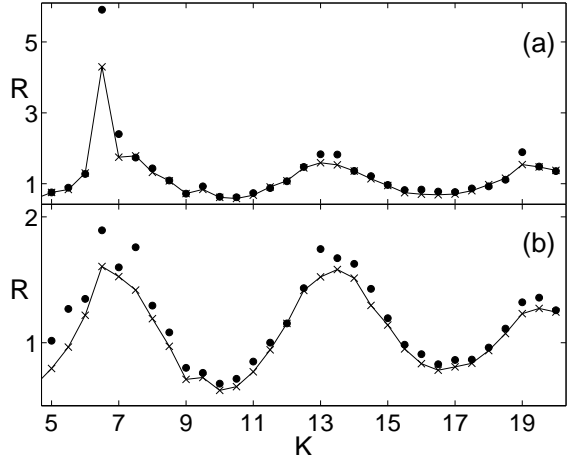}
\caption{Filled circles: The quantity $R=N^{2}(\Delta I)^{2}/(2D_{\mathrm{ql}
}) $ versus $K$ for $N=2\protect\pi /\hbar =121$ in cases $S$ [(a), $D_{
\mathrm{ql}}=K^{2}/4$] and $A$ [(b), $D_{\mathrm{ql}}=5K^{2}/4$] defined in
the caption of Fig. 1. Crosses joined by a line: $D/D_{\mathrm{ql}}$ versus $
K$.}
\label{fig2}
\end{figure}
\begin{figure}[tbp]
\includegraphics[width=7.5cm]{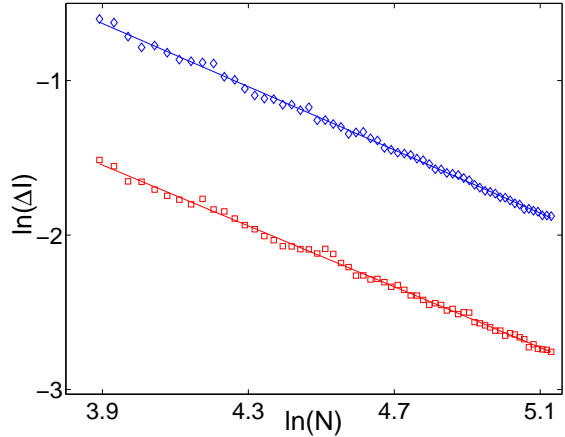}
\caption{(Color online) Loglog plots of $\Delta I$ versus $N=2\protect\pi /\hbar $ ($49\leq
N\leq 169$) for $K=15$ in case $S$ (red squares, with linear fit having slope $
\protect\mu =-0.98\pm 0.01$) and in case $A$ (blue diamonds, $\protect\mu 
=-1.02\pm 0.01$).}
\label{fig3}
\end{figure}
\begin{figure}[tbp]
\includegraphics[width=7.5cm]{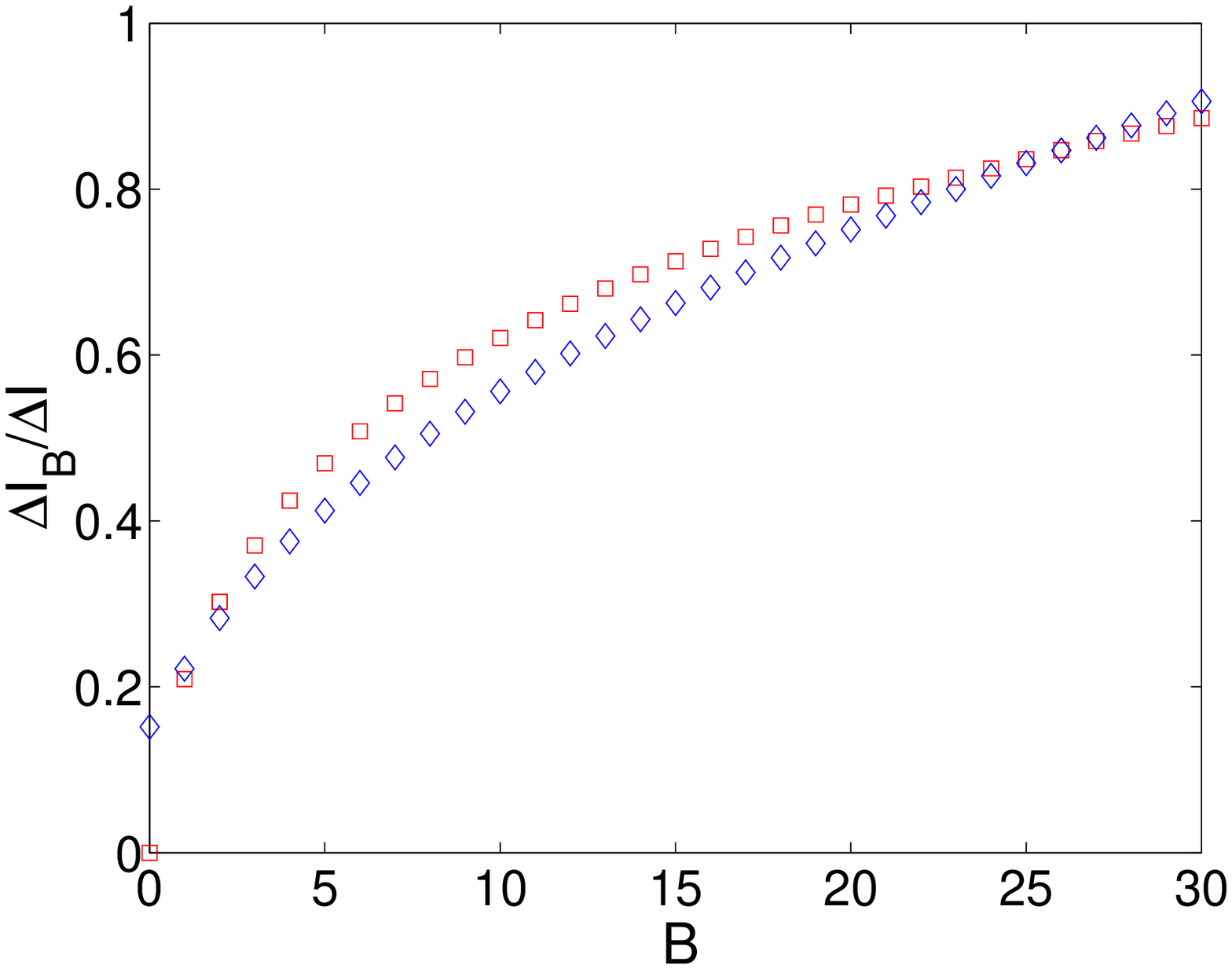}
\caption{(Color online) $\Delta I_{B}/\Delta I$ versus $B$ for $
N=121$ and $K=15$ in case $S$ (red squares) and in case $A$ (blue diamonds).}
\label{fig4}
\end{figure}

\begin{center}
{\bf V. CONCLUSIONS}
\end{center}

In conclusion, we have presented a first study of the semiclassical
full-chaos regime of the quantum ratchet effect using a novel statistical
approach which is most required in view of the sensitivity of the effect
to the initial state. For maximally uniform states $\left\vert \psi
_{\mathbf{w}}\right\rangle$ in phase space, used here for the first time
as natural initial states, the momentum-current distribution (\ref{DI})
exhibits clear fingerprints of classical chaotic diffusion, a genuine
quantum-chaos phenomenon. The simple formula for the variance in Eq.
(\ref{DI}) involves just $D$ and $\hbar^2$ and is thus interestingly
similar to the well-known one for the localization length $\xi$ in the
kicked rotor \cite{ds}, $\xi\sim D/\hbar^2$ (in our notation). This
variance was shown to be significantly larger than that for momentum
states ($B=0$ in Fig. 4), which were standardly used in previous works within the ordinary (single-state) approach. The states $\left\vert \psi
_{\mathbf{w}}\right\rangle$ then appear to give the \emph{strongest}
quantum ratchet effect known until now.\newline 

One can approximate $\left\vert\psi _{\mathbf{w}}\right\rangle $ to order $B$ by the states (\ref{kqB}) which are superpositions of $2B+1$ momentum states and whose variance increases with $B$ (see Fig. 4). Superpositions of
two momentum states were recently used in experimental realizations of
quantum-resonance ratchets \cite{qr6,qr7}. The states (\ref{kqB}) can be experimentally prepared for at
least $B\lesssim 10$ \cite{pc}. Also, the systems (\ref{khm}) are exactly
related \cite{d,d0} to kicked harmonic oscillators which are
experimentally realizable \cite{ht}. Thus, the current distributions and
the strong quantum ratchet effects predicted in this work should be
observable to some extent in the laboratory. More detailed aspects of our
statistical approach and its extension to other systems and parameter
regimes will be considered in future studies.\newline

\begin{center}
{\bf ACKNOWLEDGMENTS}
\end{center}

This work was partially supported by ISF Grant No. 118/05 and BIU Grant
No. 2046.\newline

\begin{center}
{\bf APPENDIX}
\end{center}

We derive here formula (\ref{Ipe}) and the inequality (\ref{ineq}). We start with a summary of results from Ref. \cite{d1}. Since the evolution operator $\hat{U}=\exp [-L\cos (\hat{p})/\hbar ]\exp [-KV(\hat{x})/\hbar ]$ is $2\pi $-periodic in $(\hat{x},\hat{p})$, it commutes with both $\hat{T}_{x}(2\pi )=\hat{T}_{x}^{N}(a)$ ($a=2\pi /N$) and $\hat{T}_{p}(b)$ ($b=2\pi$). Therefore, one can find simultaneous eigenstates of $\hat{U}$, $\hat{T}_{x}^{N}(a)$, and $\hat{T}_{p}(b)$. The general eigenstates of $\hat{T}_{x}^{N}(a)$ and $\hat{T}_{p}(b)$ are given in terms of (\ref{kq}) by
\begin{equation}\label{qes}
\left\vert \Psi _{j,\mathbf{w}}\right\rangle
=\sum_{m=0}^{N-1}\phi _{j}(m;\mathbf{w})\left\vert \psi
_{w_{1},w_{2}+ma}\right\rangle .
\end{equation}
Here $\phi_{j}(m;\mathbf{w})$, $j=1,\dots ,N$, form $N$ independent vectors of
coefficients, $\mathbf{V}_{j}(\mathbf{w})=\left\{ \phi _{j}(m;\mathbf{w}
)\right\} _{m=0}^{N-1}$, which are determined from the eigenvalue equation $\hat{U} \left\vert \Psi _{j,\mathbf{w}}\right\rangle =\exp [-iE_{j}(\mathbf{w}
)]\left\vert \Psi _{j,\mathbf{w}}\right\rangle $, where $E_{j}(\mathbf{w})$
are the quasienergies. It is clear from Eq. (\ref{qes}) that $\mathbf{V}_{j}(\mathbf{w})$ are the representation of $\left\vert \Psi _{j,\mathbf{w}}\right\rangle $ in the $N$-basis $\left\vert \psi _{w_{1},w_{2}+ma}\right\rangle $, $m=0,...,N-1$. In this basis, $\hat{U}$ is represented by an $N\times N$ unitary matrix $\mathbf{\hat{M}}(\mathbf{w})$ with known elements \cite{d1}. Thus, $\mathbf{\hat{M}}(\mathbf{w})\mathbf{V}_{j}(\mathbf{w})=\exp [-iE_{j}(
\mathbf{w})]\mathbf{V}_{j}(\mathbf{w})$. This completes the summary of relevant results from Ref. \cite{d1}.\newline

Let us now calculate the matrix element $\left\langle\psi _{\mathbf{w}}\vert\hat{I}\vert\psi _{\mathbf{w'}}\right\rangle$, where $\hat{I}=\lim_{t\rightarrow\infty}\hat{U}^{-t}\,\hat{p}\,\hat{U}^{t}/t$ is the momentum-current operator. First, we have
\begin{equation}\label{me}
\left\langle\psi _{\mathbf{w}}\vert\hat{U}^{-t}\,\hat{p}\,\hat{U}^{t}/t\vert\psi _{\mathbf{w'}}\right\rangle = \left\langle\psi _{\mathbf{w},t}\vert\hat{p}\vert\psi _{\mathbf{w'},t}\right\rangle /t , 
\end{equation}
where $\left\vert\psi _{\mathbf{w},t}\right\rangle =\hat{U}^{t}\left\vert\psi _{\mathbf{w}}\right\rangle$. Using the completeness of the eigenvectors $
\mathbf{V}_{j}(\mathbf{w})$ of $\mathbf{\hat{M}}(\mathbf{w})$, i.e., $\sum_{j=1}^{N}\phi _{j}^{\ast }(m;
\mathbf{w})\phi _{j}(m^{\prime };\mathbf{w})=\delta _{m,m^{\prime }}$, we can
invert Eq. (\ref{qes}) to get $\left\vert \psi _{\mathbf{w}}\right\rangle
=\sum_{j=1}^{N}\phi _{j}^{\ast }(0;\mathbf{w})\left\vert \Psi _{j,\mathbf{w}
}\right\rangle $. One then has
\begin{equation}\label{eme}
\left\langle \psi _{\mathbf{w},t}\left\vert \hat{p}
\right\vert \psi _{\mathbf{w'},t}\right\rangle =\sum_{j,j^{\prime }=1}^N\phi _{j^{\prime }}(0;\mathbf{w})\phi _{j}^{\ast }(0;
\mathbf{w'})\left\langle \hat{U}^{t}\Psi _{j^{\prime },\mathbf{w}}\left\vert 
\hat{p}\right\vert \hat{U}^{t}\Psi _{j,\mathbf{w'}}\right\rangle . 
\end{equation}
To determine the asymptotic behavior of Eq. (\ref{eme}) for large $t$, we use the equation $\hat{U}^{t}(\hat{x},\hat{p})\left\vert \Psi _{j,\mathbf{w}}\right\rangle =\exp [-itE_{j}(\mathbf{w})]\left\vert \Psi _{j,\mathbf{w}}\right\rangle $, the expansion (\ref{qes}), and the fact that $\hat{p}=-i\hbar d/dx$ is represented by $-i\hbar d/dw_{1}$ in the basis (\ref{kq}) \cite{kqz}, due to the delta comb in $x$. Using also the orthonormality of (\ref{kq}), $\left\langle\psi _{\mathbf{w}}\vert\psi _{\mathbf{w'}}\right\rangle =\delta (w_1-w'_1)\delta (w_2-w'_2)$ \cite{kqz}, and of $\mathbf{V}_{j}(\mathbf{w})$, we find that the dominant terms in $\left\langle \hat{U} ^{t}\Psi _{j^{\prime },\mathbf{w}}\left\vert \hat{p}\right\vert \hat{U}
^{t}\Psi _{j,\mathbf{w'}}\right\rangle $ for large $t$ give the asymptotic behavior
\begin{equation}\label{asy}
\left\langle \hat{U} ^{t}\Psi _{j^{\prime },\mathbf{w}}\left\vert \hat{p}\right\vert \hat{U}
^{t}\Psi _{j,\mathbf{w'}}\right\rangle \sim -t\,\hbar\frac{\partial E_{j}(\mathbf{w})}{\partial w_{1}}\delta
_{j,j^{\prime }}\delta (w_1-w'_1)\delta (w_2-w'_2), \ \ \ t\gg 1,
\end{equation}
where we assume for simplicity that $0\leq w_2,w'_2<a$. After inserting
(\ref{asy}) in (\ref{eme}) and dividing by $t$, we get from (\ref{me}) in the limit $t\rightarrow \infty $: $\left\langle\psi _{\mathbf{w}}\vert\hat{I}\vert\psi _{\mathbf{w'}}\right\rangle =I(\mathbf{w})\delta (w_1-w'_1)\delta (w_2-w'_2)$, where $I(\mathbf{w})$ is given by formula (\ref{Ipe}).\newline

To derive the inequality (\ref{ineq}), we first notice that the states (\ref{kqp}) and (\ref{kqB}) can be easily related as follows:
\begin{equation}\label{Bp}
\left\vert \psi _{\mathbf{w}}^{(B)}\right\rangle
=\int_{0}^{a}dw_{1}^{\prime }g_{B}(w_{1}^{\prime }-w_{1})\left\vert
\psi _{w_{1}^{\prime },w_{2}}\right\rangle , 
\end{equation}
where
\begin{equation}\label{gB}
g_{B}(w_{1})=\frac{1}{\sqrt{(2B+1)a}}\sum_{n=-B}^{B}\exp (2\pi inw_{1}/a).
\end{equation} 
The factor before the sum in (\ref{gB}) assures the normalization
\begin{equation}\label{norm}
\int_{0}^{a}dw_{1}g_{B}^2(w_{1})=1.
\end{equation}
Eq. (\ref{norm}) implies, because of $\left\langle\psi _{\mathbf{w}}\vert\psi _{\mathbf{w'}}\right\rangle =\delta (w_1-w'_1)\delta (w_2-w'_2)$ and $\left\langle\psi _{\mathbf{w}}\vert\hat{I}\vert\psi _{\mathbf{w'}}\right\rangle =I(\mathbf{w})\delta (w_1-w'_1)\delta (w_2-w'_2)$ (see above), that the states (\ref{Bp}) satisfy the orthonormality relation $\left\langle\psi _{w_1,w_2}^{(B)}\vert\psi _{w_1,w'_2}^{(B)}\right\rangle =\delta (w_2-w'_2)$ and $\left\langle\psi _{w_1,w_2}^{(B)}\vert\hat{I}\vert\psi _{w_1,w'_2}^{(B)}\right\rangle =I_{B}(\mathbf{w})\delta (w_2-w'_2)$, where
\begin{equation}\label{IBw}
I_{B}(\mathbf{w})=\int_{0}^{a}dw_{1}^{
\prime }g_{B}^{2}(w_{1}^{\prime }-w_{1})I(w_{1}^{\prime },w_{2}).
\end{equation}  
Then, (\ref{IBw}) is clearly the momentum current for initial state (\ref{Bp}).\newline

Now, using (\ref{norm}), (\ref{IBw}), and the Cauchy-Schwarz inequality
\[
\left\vert\int_0^a dw'_1F(w'_1)G(w'_1)\right\vert ^2\leq\int_0^a dw'_1\left\vert F(w'_1)\right\vert ^2\int_0^a dw'_1\left\vert G(w'_1)\right\vert ^2 
\]
with the identifications $F(w'_1)=g_{B}(w_{1}^{\prime }-w_{1})I(w_{1}^{\prime },w_{2})$ and $G(w'_1)=g_{B}(w_{1}^{\prime }-w_{1})$, we get:
\begin{equation}\label{cs}
I_{B}^2(\mathbf{w})\leq \int_0^a dw'_1 g_{B}^2(w_{1}^{\prime }-w_{1})I^2(w_{1}^{\prime },w_{2}). 
\end{equation}   
Using (\ref{cs}) and, again, (\ref{norm}) in the definition $\left\langle I_B^{2}\right\rangle = \int_{0}^{a}dw_{1}\int_{0}^{b}dw_{2}I_B^{2}(\mathbf{w})/h$, with $\left\langle I^{2}\right\rangle$ similarly defined, we finally obtain that $\left\langle I_B^{2}\right\rangle\leq\left\langle I^{2}\right\rangle$. This is inequality (\ref{ineq}), where $(\Delta I_{B})^2=\left\langle I_B^{2}\right\rangle$ since $\left\langle I_B\right\rangle =0$, as easily implied by (\ref{IBw}) and $\left\langle I\right\rangle =0$.\newline

\end{document}